\documentclass[twocolumn,prl,amsmath,amssymb,showpacs]{revtex4-1}

\usepackage{graphicx}
\usepackage{epstopdf}
\usepackage{hyperref}
\usepackage{url}
\usepackage{float}
\usepackage{xcolor}
\pdfoutput=1
\begin{document}
\newcommand{\ben}{\begin{equation}}
\newcommand{\een}{\end{equation}}


\title{Magnetic Domain Wall Floating on a Spin Superfluid}

\author{Pramey Upadhyaya}  
\affiliation{Department of Physics and Astronomy, University of California, Los Angeles, California 90095, USA}

\author{Se Kwon Kim}  
\affiliation{Department of Physics and Astronomy, University of California, Los Angeles, California 90095, USA}

\author{Yaroslav Tserkovnyak}
\affiliation{Department of Physics and Astronomy, University of California, Los Angeles, California 90095, USA}

\begin{abstract}
We theoretically investigate the transfer of angular momentum between a spin superfluid and a domain wall in an exchange coupled easy-axis and easy-plane magnetic insulator system. A domain wall in the easy-axis magnet absorbs spin angular momentum via disrupting the flow of a superfluid spin current in the easy-plane magnet. Focusing on an open geometry, where the spin current is injected electrically via a nonequilibrium spin accumulation, we derive analytical expressions for the resultant superfluid-mediated motion of the domain wall. The analytical results are supported by micromagnetic simulations. The proposed phenomenon extends the regime of magnon-driven domain-wall motion to the case when the magnons are condensed and exhibit superfluidity. Furthermore, by controlling the pinning of the domain wall, we propose a realization of a reconfigurable spin transistor. The long-distance dissipationless character of spin superfluids can thus be exploited for manipulating soliton-based memory and logic devices.

\end{abstract}


\maketitle

\textit{Introduction.}|Spin currents carried by collective excitation of magnets, in lieu of charge currents, have recently attracted vibrant experimental and theoretical activities opening a subfield of spintronics dubbed \textit{magnonics} \cite{Chumak2015}. This is motivated in part by the prospects of constructing low-dissipation spintronic devices. Apart from allowing for the Joule heating-free transfer of spin signals, magnons also offer the possibility of imparting their spin angular momentum to topological solitons \cite{PhysRevLett.107.027205, *PhysRevLett.107.177207, *alexey_dw, *PhysRevLett.111.067203, *PhysRevLett.111.067203, *PhysRevB.89.064412, *PhysRevB.89.241101, *PhysRevB.90.094423}. These solitons \cite{Kosevich1990}, such as domain walls and skyrmions, are robust against fluctuations and are thus considered as ideal candidates for encoding nonvolatile information \cite{Parkin11042008, *Nagaosa2013, *skyrmionics}. Recent experimental demonstrations of thermal magnon-induced domain-wall \cite{PhysRevLett.110.177202} and skyrmion motion \cite{Mochizuki2014} could thus provide a basis for all-magnonic nonvolatile memory (such as the race-track register \cite{Parkin11042008}) and logic devices \cite{Allwood2005}.

On another front, these magnons offer a unique possibility of forming coherent condensates at room temperature, as demonstrated experimentally by parametric (microwave) pumping in a magnetic insulator \cite{Demokritov2006}. Such condensates present an exciting opportunity for magnonics by supporting a long-distance coherent superfluid-like transport of the spin current \cite{Sonin2012, *2015arXiv150300482B}, as opposed to the exponentially decaying  spin currents carried by the incoherent thermal magnons. In addition to the pumped systems, such spin superfluidity is also supported by easy-plane magnets having a $U(1)$ order parameter \cite{Sonin2010}. More recently, these spin superfluids are gaining increased attention with proposals of realizing them in various easy-plane systems \cite{PhysRevLett.87.187202, *PhysRevLett.112.227201, *PhysRevB.90.220401, *PhysRevLett.115.237201, *PhysRevLett.115.156604, *PhysRevB.89.024511, *2015arXiv150601061T} \cite{PhysRevB.90.094408}. The superfluid nature of spin currents results in: an algebraically decaying transport of spin \cite{PhysRevLett.112.227201}, magnetic analogues of the Josephson effect \cite{PhysRevB.90.144419, *PhysRevB.89.024511}, dissipation via phase slips \cite{PhysRevB.93.020402, *PhysRevLett.116.127201}, and macroscopic qubit functionality \cite{2015arXiv151204546T}. While these proposals establish the feasibility of an efficient transport of the spin information, the possibility of transferring angular momentum by these superfluid-like spin currents remains unexplored. In this Letter, we fill this gap by proposing a scheme for coupling spin currents carried by superfluids to magnetic solitons.

\begin{figure}[t!]
\centering
\includegraphics[width=0.5\textwidth, scale=0.9]{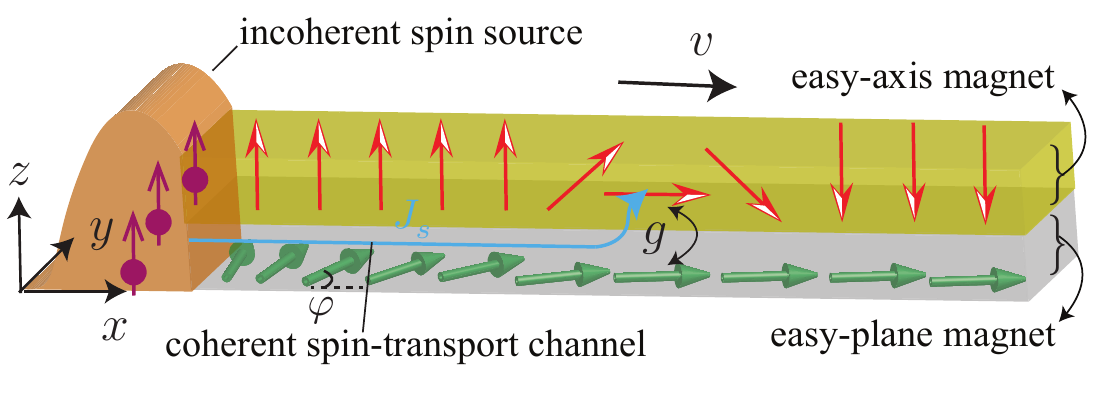}
\caption{A bilayer of an easy-$z$-axis magnet exchange coupled (with the coupling strength $g$) to an easy-$xy$-plane magnet. A $z$-polarized spin current is injected from an incoherent spin source and propagates as a superfluid spin current through the easy-plane magnet. This spin current is $\propto \nabla \varphi$, where $\varphi$ is the azimuthal angle of the spin order parameter within the $xy$ plane. This spin current is interrupted and absorbed by a domain wall in the easy-axis magnet, where it is converted into its sliding motion at speed $v$.}
\label{schematic_main}
\end{figure}

The main idea is to form an exchange coupled bilayer of an easy-plane and an easy-axis magnetic insulator. The bilayer is rotationally invariant about an axis of symmetry, which coincides with the easy axis and the normal to the easy plane. See Fig.~\ref{schematic_main} for a schematic (where $z$ is the symmetry axis). The easy-plane magnet plays the role of a spin superfluid and the easy-axis magnet harbors a domain wall. When a spin current polarized along the symmetry axis is injected into the bilayer, it is transported coherently by the gradient of the azimuthal angle ($\varphi$) of the spin density in the easy-plane magnet \cite{Sonin2010}. A static domain wall blocks the flow of this spin current by pinning $\varphi$ underneath the domain wall. The pinning occurs due to the finite exchange coupling between the spin densities in the easy-axis and the easy-plane magnets. However, the $U(1)$ symmetry of the combined system demands conservation of the total spin current polarized along the symmetry axis. Consequently, the coherently transported spin current in the easy-plane magnet is absorbed by the domain wall and converted into its motion. The problem of deriving analytical expressions for this spin transfer-induced domain-wall motion and using it to propose a spin transistor are the main focus of this Letter. Our proposal extends the concept of magnon-induced torques (due to the exponentially decaying incoherent magnons \footnote{At a finite temperature, angular momentum can also be transported by an incoherent channel of thermal magnons within the easy-axis magnet itself \cite{Cornelissen2015}. These magnons can also apply torque on the domain wall.}) to the more efficient case, where the magnons are condensed and exhibit superfluidity.


\textit{Model.}|We focus on a one-dimensional model with a bilayer strip extended along the $x$ axis. The free energy (density) of the system can then be written as:
\ben
\mathcal{F}= A |\partial_x {\bf m}|^2/2-Km_z^2/2+ \mathcal{F}_{\rm sf}+U_{\rm int},
\label{free_en_bilayer}
\een
where $A$, $K>0$ and ${\bf m}$ represent the magnetic stiffness, the anisotropy and the unit vector oriented along the spin density in the easy-axis magnet, respectively. $\mathcal{F}_{\rm sf}$ is the free energy of the spin superfluid and $U_{\rm int}$ is the exchange coupling-induced interaction between the easy-axis and the easy-plane magnets. Within the easy-axis magnet, the equilibrium configuration of interest is that of magnetic domains (referred to as regions I and III for ${\bf m}$ along $+{\bf z}$ and $- {\bf z}$, respectively) separated by a single domain wall (referred to as the region II). 

\begin{figure}
\centering
\includegraphics[width=0.5\textwidth, scale=0.8]{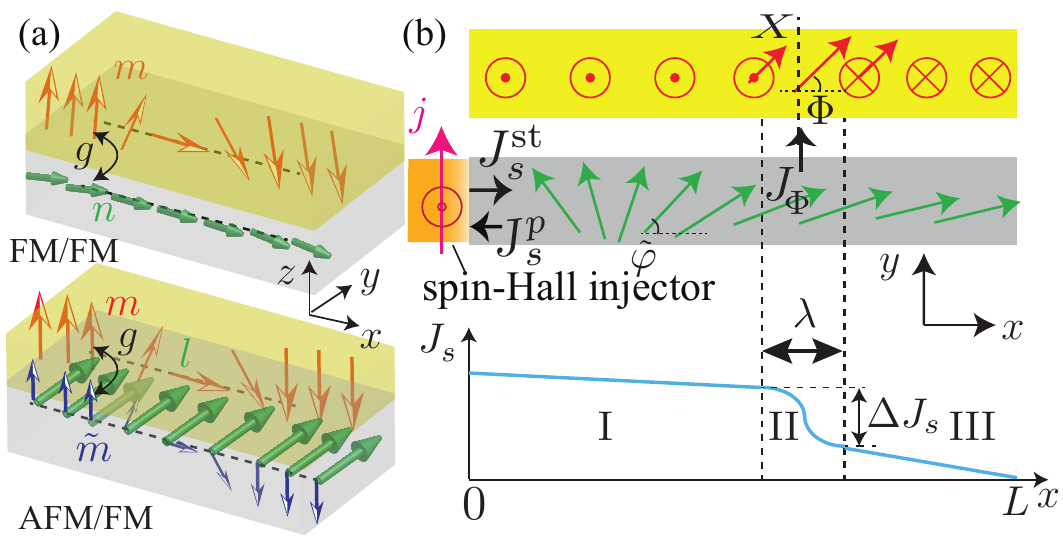}
\caption{(a) Equilibrium configurations for exchange coupling the spin densities of an easy-axis ferromagnet, ${\bf m}$, with: (top panel) an  easy-plane ferromagnet, ${\bf n}$, and (bottom panel) the net spin density ${\bf \tilde{m}}$ (resulting from canting of the sublattice spin densities) of a Heisenberg antiferromagnet. Here, the N{\'e}el vector is denoted by ${\bf l}$. (b) Top panel: the model of a domain wall of width $\lambda$ coupled to a spin superfluid. The domain wall divides the bilayer into three regions: up domain (I), down domain (III) and the domain wall (II). A spin current, $J_s^{\rm st}$, is injected on the left by converting a charge current, $j$, into a spin accumulation via the spin Hall effect. Upon reaching the domain-wall region a portion of this spin current, $J_\Phi$, is absorbed from the easy-plane magnet by the domain wall. The resultant dynamics of the domain wall is characterized by the generalized coordinates $X$ and $\Phi$, parametrizing its position and the associated azimuthal angle. The dynamics of the spin superfluid pumps a spin current, $J_s^p$, back to the contact. Bottom panel: the corresponding superfluid spin current flowing in the easy-plane magnet.}
\label{schematic_dw_motion}
\end{figure}

We discuss two possible routes for forming the proposed bilayer. That is, when the spin superfluid is (1) an easy-plane ferromagnet [labeled as FM/FM in Fig.~\ref{schematic_dw_motion} (a)], or (2) a Heisenberg antiferromagnet [labeled as AFM/FM in Fig.~\ref{schematic_dw_motion} (a)]. For the FM/FM case, $\mathcal{F}_{\rm sf}=\tilde{A}|\partial_x {\bf n}|^2/2 + \tilde{K}n_z^2/2$ and $U_{\rm int}=U_{\rm int}^{F}=-g {\bf m} \cdot {\bf n}$. Here, $\tilde{A}$ and ${\bf n}$ are the magnetic stiffness and the unit vector aligned with the spin density in the easy-plane ferromagnet, respectively, and $g$ is the strength of the exchange coupling. The easy-plane character is enforced by having  $\tilde{K} >0$ \footnote{We focus on the regime where $g/K$ and $g/\tilde{K} \ll 1$. Within this regime, deviations from the desired equilibrium configuration of a domain wall in the easy-axis magnet and $\bf n$ lying within the easy plane are small}. See Fig.~\ref{schematic_dw_motion} (a), top panel, for a schematic of the equilibrium configuration. In contrast to the FM/FM case, a natural easy-plane system is formed in the AFM/FM case, where $\mathcal{F}_{\rm sf}=\tilde{A}|\partial_x {\bf l}|^2/2 + {\bf \tilde{m}}^2/2\chi$ and $U_{\rm int}=U_{\rm int}^{\rm AF}=-g {\bf m} \cdot {\bf \tilde{m}}$. Here, ${\bf l} \equiv ({\bf \tilde{m}}_1 -{\bf \tilde{m}}_2)/2$ and ${\bf \tilde{m}} \equiv {\bf \tilde{m}}_1 +{\bf \tilde{m}}_2$ are vectors oriented along the staggered and net spin densities, respectively, with ${\bf m}_1$ and ${\bf m}_2$ being the unit vectors along the sublattice spin densities of an isotropic antiferromagnet. In order to minimize $U_{\rm int}$, ${\bf \tilde{m}}$ follows ${\bf m}$. The orthogonality of ${\bf l}$ and ${\bf \tilde{m}}$ then ensures ${\bf l}$ to lie in the $xy$ plane within regions I and III. See Fig.~\ref{schematic_dw_motion} (a), bottom panel, for a schematic of the equilibrium configuration. The gradient of ${\bf l}$ can then transport the superfluid spin current \cite{PhysRevB.90.094408}.

\textit{Coupled spin hydrodynamics.}|We begin by outlining a hydrodynamic theory for describing the proposed spin superfluid mediated domain-wall motion. The central idea is to write down the continuity equation for the flow of the $z$ component of the spin current in the bilayer. See Fig.~\ref{schematic_dw_motion} (b) for a schematic. In regions I and III  this spin current is transported within the easy-plane magnet. In the strong anisotropy and the long-wavelength limit of the spin dynamics, the transport is described by \cite{PhysRevLett.112.227201} \cite{PhysRevB.90.094408}:
\ben
\tilde{s}\dot{\tilde {n}}_z=\tilde{A}_t\partial_x^2\tilde{\varphi}- \tilde{s}\tilde{\alpha} \dot{\tilde{\varphi}},
\label{super_coupling}
\een 
where $\tilde{s}$ is the magnitude of the saturated spin density per unit area (i.e. integrated over the thickness $t_p$ of the easy-plane magnet). Here, we have defined $\tilde{A}_t \equiv \tilde{A} t_p$, while $\tilde{n}_z$ and $\tilde{\varphi}$ represent respectively the $z$ component and the azimuthal angle of the unit vector oriented along the spin order parameter in the easy-plane magnet. For the FM/FM and the AFM/FM case, this spin order parameter is given by ${\bf n}$ and ${\bf l}$, respectively. The first term on the right-hand side defines a superfluid spin current $J_s \equiv -A_t\partial_x \tilde{\varphi}$, and the second term describes the transfer of the spin current to the atomic lattice due to a finite Gilbert damping, $\tilde{\alpha}$, within the easy-plane magnet. In region II, additional spin current, $J_\Phi$, is absorbed by the domain wall. Using the collective coordinate approach \cite{Thiele, *Bazaliy2008}, the resultant domain-wall dynamics can be written as:
\begin{subequations}
\begin{align}
2s\dot{\Phi} + 2\alpha s \dot{X}/\lambda &= 0 \label{force}  \\
2s \dot{X}- 2\alpha s \lambda \dot{\Phi} &= J_\Phi,
\label{torque}
\end{align}
\label{DW_EOM}
\end{subequations}
where the so-called \textit{soft} modes $X$ and $\Phi$ represent the location at which the $z$ component of the spin density vanishes in the wall and the azimuthal angle at this location, respectively. Here, $\lambda$ is the domain-wall width and $s$ is the magnitude of the saturated spin density (integrated over the the thickness $t_e$ of the easy-axis magnet).
Eq.~(\ref{torque}) describes the flow of the spin current within the domain-wall region. Namely, the spin current absorbed by the domain wall is converted into its motion, giving rise to the term proportional to $\dot{X}$. In addition, a portion of the absorbed spin current is transferred to the atomic lattice in the easy-axis magnet, resulting in the term proportional to $\alpha$. 

In the spirit of the long-wavelength spin dynamics, throughout this Letter, we consider the domain wall as a point-like object satisfying $\lambda \sim \sqrt{A/K} \ll 1/\partial_x\tilde{\varphi}$. In this case, the width of the region II can be neglected, and the discontinuity in the spin current flowing in the easy-plane magnet at $x=X$ [see the bottom panel of Fig.~\ref{schematic_dw_motion}(b)] is given by: $|\Delta J_s|=J_\Phi=2s(1+\alpha^2)\dot{X}.$
Equipped with this boundary condition, at $x=X$, we are now ready to discuss the motion of the domain wall in response to a spin current injected from the left of the bilayer. For this purpose, we consider the open geometry proposed in Ref. \onlinecite{PhysRevLett.112.227201}, whereby we solve  Eqs.~(\ref{super_coupling}) and (\ref{DW_EOM}) subject to the following additional boundary conditions: $J_s|_{x=0}=J_s^{\rm st}-J_s^p$ and $J_s|_{x=L}=0$. The former corresponds to the injection of a spin current from a metallic contact using the spin-Hall effect \cite{PhysRevLett.83.1834}, while the latter condition is equivalent to the usual exchange boundary condition at the right boundary. Within spin Hall phenomenology \cite{PhysRevB.90.014428}, $J_s^{\rm st} \equiv \vartheta j$, with $j$ being the charge current density (per unit length) at the metal/easy-plane magnet interface [see Fig.~\ref{schematic_dw_motion}(b)], and $\vartheta \equiv \hbar t_p \tan \theta/ 2el_N$. Here, $\theta$, $e$, and $l_N$ denote the so-called spin Hall angle \cite{6516040}, charge of an electron, and length (along the $x$ axis) of the metallic contact, respectively. $J_s^p \equiv \gamma^{\uparrow \downarrow}n \times \dot{n}$ represents the spin current pumped back in the left contact, with $\gamma^{\uparrow \downarrow} \equiv\hbar t_p g^{\uparrow \downarrow}/4\pi$ \cite{PhysRevLett.88.117601}. Here, $g^{\uparrow \downarrow}$ parametrizes the real part of the spin mixing conductance for the metallic contact/easy-plane magnet interface.

\textit{Superfluid-induced domain-wall motion: Linear regime.}|We proceed to look for solutions of the form $\dot{\Phi}=\Omega$,  $\varphi(x,t)=f(x)+\Omega t$ and $\dot{n}_z=0$. Physically, such an ansatz represents a linearly decaying spin current in regions I and III \cite{PhysRevLett.112.227201}, and a steady-state motion of the domain wall, with $\dot{X}=v$. We highlight that within this ansatz, the domain-wall angle is preccessing at the same frequency as the underlying spin superfluid and refer to this dynamic regime as the ``locked" phase. Furthermore, in the presence of a moving domain wall, the assumption of having a position independent $\Omega$ is not self evident. We justify and discuss its validity \textit{a posteriori} \footnote{See Supplementary materials for the condition of validity}. Balancing the flow of spin current, via substitution of the ansatz in Eqs.~(\ref{super_coupling}), (\ref{DW_EOM}) and the boundary conditions, yields \footnote{we have retained leading order terms in damping, neglecting terms of the form $\sim \alpha^2$, $\alpha \tilde{\alpha}$. However, $\alpha^2 L/\lambda$ is kept since $L/\lambda \gg 1$.}:
\begin{align}
v &= \frac{\vartheta j}{2s+\alpha(\gamma^{\uparrow \downarrow}+\gamma_\alpha)/\lambda}.
\label{sf_vel}
\end{align}
Here, we have used ${\bf n}\times \dot{{\bf n}}= \Omega {\bf z}$ and defined $\gamma_\alpha \equiv \tilde{\alpha} \tilde{s} L$. This is one of the central result of the model describing superfluid-induced velocity of the domain wall. In the absence of the Gilbert damping, all of the injected spin current is absorbed by the domain wall giving a velocity obtained by the conservation of the angular momentum, i.e. $ v=\vartheta j/2s$. While, the loss of the spin current results in a reduction of the velocity from this perfect absorption case. Similar to the case of transport of superfluid spin current \cite{PhysRevLett.112.227201}, this loss of spin current has two sources: (a) interfacial (due to spin-pumping), giving rise to the term proportional to $\gamma^{\uparrow \downarrow}$, and (b) bulk, giving an algebraically decaying velocity with the length of the bilayer.

\textit{Superfluid-induced domain-wall motion: Nonlinear regime.}|For a critical strength of the external drive, the steady-state ansatz of the domain wall moving with a linearly increasing velocity breaks down. This phenomenon is referred to as the Walker breakdown \cite{walker1974} and is observed for both external field and current-induced domain-wall motion \cite{Beach2008}. In this section we focus on the analogue of the Walker breakdown phenomenon for the superfluid-mediated spin transfer. For this purpose, we derive an analytical expression of $J_\Phi$ within the Landau-Lifshitz phenomenology. The $z$ component of the torque applied on the easy-axis magnet, due to the coupling to the easy-plane magnet, reads as: $\tau_z=-{\bf z} \cdot {\bf m} \times \delta_{\bf m} U_{\rm int}$. The spin current absorbed by the domain wall is then given by integrating the torque over the domain-wall region, i.e. $J_\Phi= t_p \int_\lambda \tau_z dx$. We adopt the following parametrization of the Cartesian components of the unit vector field: ${\bf m} \equiv (\sin \theta \cos \phi, \sin \theta \sin \phi, \cos \theta)$. Here, $\theta$ and $\phi$ are respectively the polar and the azimuthal angles of the spin density in the easy-axis magnet. For the FM/FM case, substituting the parametrization ${\bf n} \equiv (\cos \varphi, \sin \varphi, n_z)$ in $U_{\rm int}^F$, we get (up to linear order in $n_z$) $J_\Phi^F=\pi g \lambda t_p \sin(\Phi-\varphi|_X)$. Here, $\varphi|_X$ is the value of $\varphi$ at $X$. Similarly, for the AFM/FM system, substituting ${\bf \tilde{m}} \sim \chi g {\bf l \times m \times l}$ and ${\bf l} \equiv (\cos \varphi, \sin \varphi, l_z)$ in $U_{\rm int}^{\rm AF}$ yields $J_\Phi^{\rm AF}=\chi g^2 t_p \lambda \sin[2(\Phi-\varphi|_X)]$. Here, we have neglected the higher order dynamic corrections to ${\bf \tilde{m}}$, which is assumed to take on its equilibrium value obtained by minimizing $\mathcal{F}$ under the constraint ${\bf l \cdot \tilde{m}}=0$.

\begin{figure*}[t!]
\centering
\includegraphics[width=\textwidth, scale=0.95]{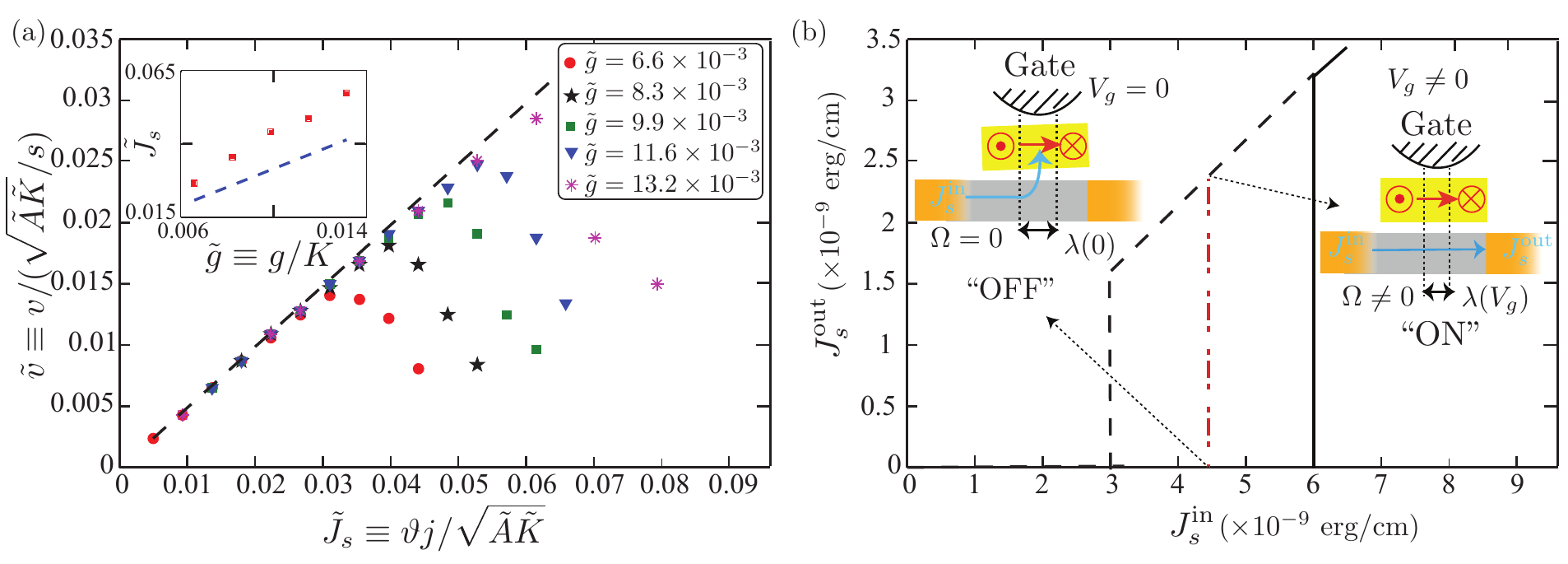}
\caption{(a) For a given exchange coupling $\tilde{g}$, two regimes for domain-wall motion are obtained. A steady-state regime with linearly increasing velocity ($\tilde{v}$) and oscillatory motion above a critical value of injected spin current $\tilde{J}_s$. Broken line plots the analytical result from Eq.~(\ref{sf_vel}). Inset shows that the critical $\tilde{J}_s$ increases linearly with $\tilde{g}$. Broken line shows the analytical result from Eq.~(\ref{critical}). (b) The spin current detected at the right end of the bilayer ($J_s^{\rm out}$) exhibits a nonlinear behavior in the presence of a pinned domain wall. When the injected spin current, $J_s^{\rm in}$, is below (above) a critical breakdown current,  $J_s^{\rm out}=0$ ($J_s^{\rm out} \neq 0$). Solid and broken curves plot this non-linear characteristics for $\lambda=10$~nm and $\lambda=5$~nm, respectively. The nonlinearity can be used to construct a transistor, as indicated by the vertical dash-dot line. Fixing $J_s^{\rm in}$ and changing $\lambda$ by an external gate switches the device from an OFF ($J_s^{\rm out}=0$) to an ON ($J_s^{\rm out} \neq 0$) state. These OFF and ON states are depicted schematically in the insets.}
\label{numerical}
\end{figure*}

For a given coupling $g$, there exists a maximum value of the absorbed spin current $J_\Phi^c$, i.e. when $\Phi-\varphi|_X=\pi/2$ for the FM/FM and $\Phi-\varphi|_X=\pi/4$ for the AFM/FM system. This results in a corresponding critical value for the injected spin current, $J_c^s$, and a critical domain wall velocity [from Eq.~(\ref{torque})], $v_c \sim J_c^s/2s$, above which the locked phase can no longer exist. Namely, $\Phi$ and $\varphi|_X$ precess at different frequencies, resulting in an oscillatory exchange of the spin current between the domain wall and the spin superfluid. We refer to this transition as a locked to unlocked breakdown. Consequently, as in the case of the Walker breakdown, the domain wall is expected to drift in an oscillatory fashion, with $\langle v\rangle <v_c$. Substituting the value of critical velocity in Eq.~(\ref{sf_vel}), we obtain for the breakdown spin current:
\ben
J_c^s= \vartheta j_c = \eta g \lambda t_p \left[1+\frac{\alpha(\gamma^{\uparrow \downarrow}+ \gamma_\alpha)}{2s\lambda}\right],
\label{critical}
\een
with $\eta=\pi$ for the FM/FM, while $\eta=\chi g$ for the AFM/FM case. This is the second main result of the model, predicting a linear dependence of the breakdown spin current on $\lambda$. Note that the transition from the locked phase to the unlocked phase is analogous to the transition of superconducting Josephson junctions from zero-voltage state to finite-voltage state \cite{tinkham2004introduction}. Below, we exploit the dependence of the critical (injected) spin current on the domain-wall width for proposing a spin transistor.

In Fig.~\ref{numerical}(a), we compare the analytical results with micromagnetic simulations \footnote{See Supplementary materials for details of the micromagnetic simulations}. As predicted by the model, two regimes are observed in the simulations: (a) linearly increasing domain-wall velocity below a critical value of the injected spin current ($J_s^c$), and (b) oscillatory drift of the domain wall with a reduced average velocity above $J_s^c$. Moreover, both the velocity in the linear regime and the value of the critical current for locked to unlocked breakdown agrees well with the simulations.

\textit{Spin transistor.}|We propose to utilize the domain-wall width dependence of the locked to unlocked breakdown in conjunction with the voltage control of the magnetic anisotropy (VCMA) \cite{maruyamaNATNANO09} to construct a spin transistor. For this purpose we consider the case of a strongly pinned domain wall, i.e. with $\dot{X}=\dot{\Phi}=0$. The pinning of $\Phi$ could be achieved by fabricating a nanowire geometry for the easy-axis magnet. In this case, the dipolar interaction favors $\Phi$, such that, the domain-wall magnetization is oriented along the long axis of the nanowire. The domain-wall position can be pinned by engineering ``notches", which create a local energy minima with respect to $X$ \cite{Parkin11042008}. For an injected spin current $J_s^{\rm in}\equiv \vartheta j < J_\Phi^c$, a static solution results for the spin superfluid with the domain wall absorbing all of the spin current injected at the left contact. See the ``OFF" schematic in the inset of Fig.~\ref{numerical}(b). Consequently, for $J_s^{\rm in} < J_\Phi^c$, a detector of the spin current placed at the right boundary would register zero spin current. On the other hand, for $J_s^{\rm in} >J_\Phi^c$ locked to unlocked breakdown occurs, resulting in a precessing solution for the superfluid. Since $J_\Phi \propto \sin(\Phi-\varphi_X)$, the spin current absorbed by the domain wall averages to zero. Utilizing the inverse spin Hall effect \cite{ISHE}, the spin current beyond the domain wall can be detected by adding a right metal contact. See the ``ON" schematic in the inset of Fig.~\ref{numerical}(b). Focusing on the case when the interfaces dominate over the bulk, i.e. $\gamma^{\uparrow \downarrow} \gg \gamma_\alpha$, half of the spin current is pumped back to the left contact and the other half is detected by the right contact, i.e. $J_s^{\rm out}=J_s^{\rm in}/2$. Here, the interfaces are assumed to be symmetric, parametrized by the same $\gamma^{\uparrow \downarrow}$. The $\lambda$ dependence of $J_\Phi^c$ then translates into the following transistor-like action [plotted in Fig.~\ref{numerical}(b)]. The ``OFF" (``ON") state of the device is defined as $J_s^{\rm out}$ being zero (nonzero). In the absence of the gate voltage, $V_g$, the device is biased to be below the locked to unlocked breakdown and hence in the OFF-state. Application of a gate voltage changes $\lambda$ (by changing $K$ via VCMA) and turns the device ON abruptly, via inducing locked to unlocked breakdown. We note that, the proposed spin transistor has an added advantage. Namely, the domain wall can be moved to a desired location by applying a magnetic field, making the device reconfigurable.  

In summary, we have proposed using spin superfluids as interconnects for transferring spin angular momentum to solitons in a coherent fashion. 
Although we have focused on the case of one dimensional domain walls, similar phenomenon will occur for two-dimensional textures, such as skyrmions. Additionally, we note that a mechanism to amplify the spin current is needed for concatenating the proposed spin transistors. In this regard, the Onsager reciprocal process of superfluid-induced domain-wall motion may be utilized. Namely, the motion of the domain wall in the easy-axis magnet (via application of an external magnetic field) would pump spin current in the easy-plane magnet. 

\begin{acknowledgments}
This work was supported by FAME (an SRC STARnet center sponsored by MARCO and DARPA). 
\end{acknowledgments}

\end{document}


\newcommand{\ben}{\begin{equation}}
\newcommand{\een}{\end{equation}}


\title{Supplemental Material: Magnetic Domain Wall Floating on a Spin Superfluid}

\author{Pramey Upadhyaya}  
\affiliation{Department of Physics and Astronomy, University of California, Los Angeles, California 90095, USA}

\author{Se Kwon Kim}  
\affiliation{Department of Physics and Astronomy, University of California, Los Angeles, California 90095, USA}

\author{Yaroslav Tserkovnyak}
\affiliation{Department of Physics and Astronomy, University of California, Los Angeles, California 90095, USA}

\maketitle

In this supplemental material, we provide the details of the micromagnetic simulations used to demonstrate the analytical results of (a) superfluid-induced steady state motion of the domain wall, and (b) the locked to unlocked breakdown. Additionally, we provide the self-consistency condition for the validity of the assumed uniform precession ansatz for the linear regime of domain-wall motion in the main text.

\textit{Micromagnetics.}|We use the LLG Micromagnetic simulator \cite{LLG} to simulate coupled one-dimensional spin chains with a free energy as described by Eq.~(1) in the main text. The dynamics of the coupled fields is obtained by integrating the Landau-Lifshitz-Gilbert equations: $s(1+\alpha {\bf m} \times)\partial_t {\bf m}=-{\bf m} \times \delta_{\bf m} \mathcal{F}$ and $\tilde{s}(1+\tilde{\alpha} {\bf n} \times)\partial_t {\bf n}=-{\bf n} \times \delta_{\bf n} \mathcal{F}$. Using parameters for typical perpendicular magnets \cite{Ikeda2010}, we take $M_s=800$~emu/cc and $\gamma= 1.7 \times 10^11$~rad/s$\cdot$Tesla. The saturated spin density is then given by $s=\tilde s= M_s/\gamma$, with $M_s$ and $\gamma$ being the saturation magnetization and the gyromagnetic ratio, respectively. The exchange constant and the anisotropies are chosen to be $A=\tilde{A}=1\mu$erg/cm and $K=\tilde{K}$, such that the domain-wall width $\lambda \sim \sqrt{A/K}= 10$~nm, while the coupling $g$ is varied as indicated in Fig.~3 (a) of the main text. The size of the system was fixed to be sufficiently long, i.e. $L \sim 100\lambda$. The size of the unit cell, after discretizing the system, was varied from $5$~nm to $2$~nm. The results were checked to be independent of the discretization size. Additionally, all simulations were performed at zero temperature and with a damping parameter of $\alpha=\tilde{\alpha}=0.01$.

\begin{figure}[h!]
\centering
\includegraphics[width=0.5\textwidth]{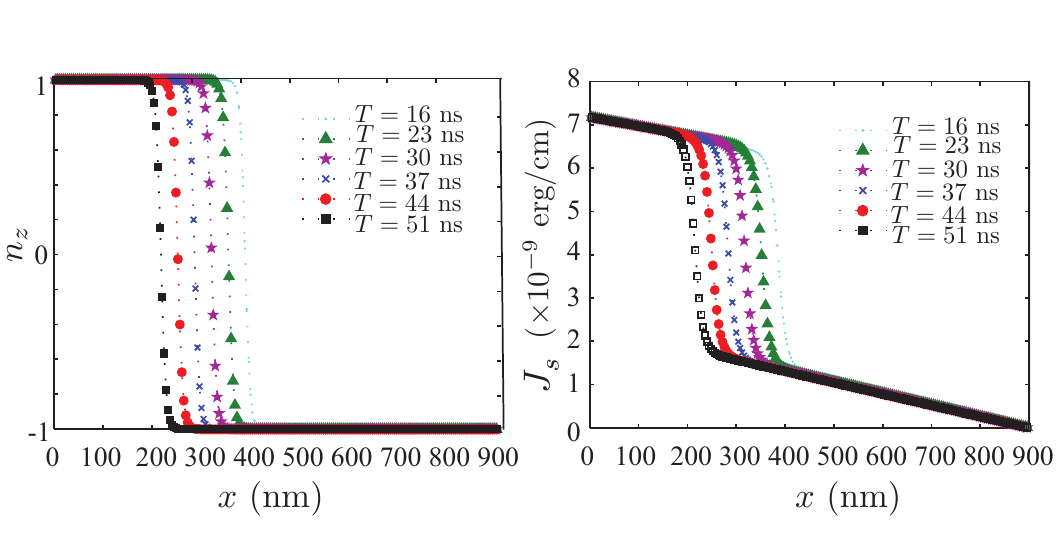}
\caption{The linear regime of superfluid-induced domain wall motion. Left panel: the out of plane component of magnetization ($n_z$) in the easy-axis magnet as a function of time $T$ after the injection of spin current at the leftmost site. The injected spin current $J_s^{\rm in} \sim 7 \times 10^{-9} < J_s^c$. The domain wall position (location at which $n_z=0$) drifts with a constant velocity. Right panel: The superfluid-like spin current, $J_s = -A_t\partial_x \phi$, flowing within the easy-plane magnet as  extracted from the micromagnetic simulations. In steady state, this spin current falls linearly within the domains, while a constant spin current is absorbed at the domain-wall location.}
\label{micro1}
\end{figure}

\begin{figure}[h!]
\centering
\includegraphics[width=0.5\textwidth]{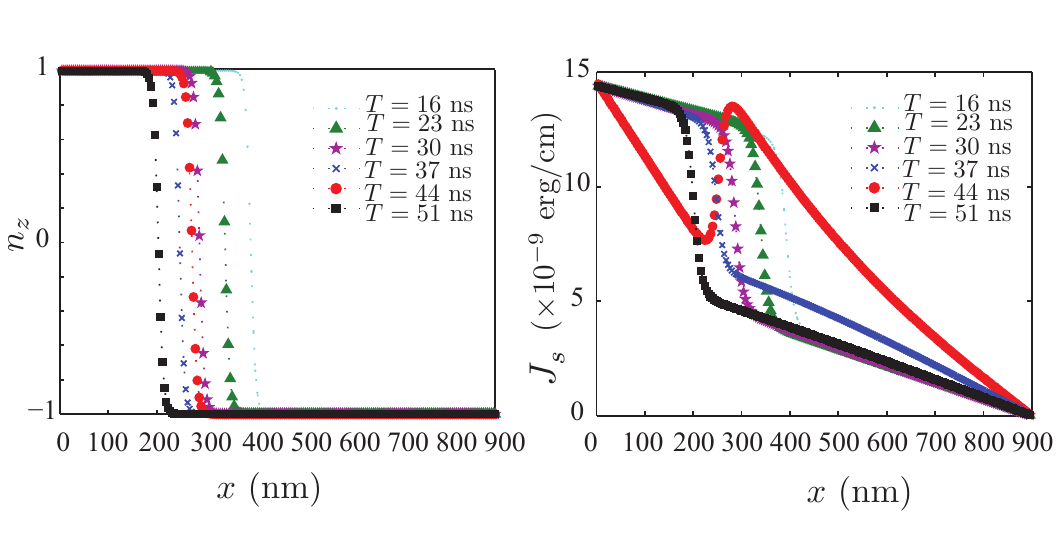}
\caption{Nonlinear regime of superfluid-induced domain wall motion. Left panel: The domain wall drifts in an oscillatory manner for injected spin current $J_s^{\rm in} \sim 15 \times 10^{-9} > J_s^c$. Right panel: Corresponding superfluid-like spin current within the easy plane showing oscillatory exchange of spin angular momentum between the domain wall and the spin superfluid.}
\label{micro2}
\end{figure}

First, an equilibrium configuration was obtained by relaxing the magnetic configuration from the initial configuration of a single domain wall in the easy-axis magnet and a uniform magnetization (oriented along the $x$ axis) for the easy-plane magnet. Next, a nonzero spin current of varying magnitude was injected for the leftmost site, by adding a damping-like spin orbit torque to the coupled dynamics given by $\vartheta j {\bf n}_l \times {\bf n}_l \times {\bf z}$. Here, ${\bf n}_l$ is the spin density unit vector for the left most site of the easy-plane magnet. In Figs.~\ref{micro1} and ~\ref{micro2}, we show the response of the domain wall, and the corresponding spin current absorbed by it, as a result of this procedure. The critical breakdown current for the chosen bilayer parameters was $J_c^s \sim 12 \times 10^{-9}$~erg/cm. After initial transients (not shown), the system settles down into a steady state. For $\vartheta j < j_c$ (Fig.~\ref{micro1}), the domain wall moves with a constant velocity, while for $\vartheta j > j_c$ (Fig.~\ref{micro2}) the domain wall drifts in an oscillatory fashion. In the former case, a constant spin current is absorbed by the domain wall, while in the latter case the spin current absorbed by the domain wall oscillates between a positive and a negative (implying emission of spin current into the easy-plane magnet) value. In Fig.~3 (a) of the main text we plot the average domain-wall velocity in the steady state.

\textit{Validity condition of uniform ansatz.}|For internal consistency, we check here for the validity of the assumed ansatz of a uniform precession frequency $\Omega$, within the linear regime of domain-wall motion. Noting  $\partial_x \partial_t \varphi=\partial_t \partial_x \varphi=-\partial_t J_s/\tilde{A}_t$, within our obtained solution for the spin current, a moving domain wall results in a difference in the precession frequency between regions I and  III: $|\Delta \Omega|= v|\Delta J_s|/\tilde{A}_t=vJ_\Phi/\tilde{A}_t \sim 2sv^2/\tilde{A}_t$. Being a quadratic correction, $\Delta \Omega$ can be neglected within the linear response. Moreover, comparing $|\Delta \Omega|$ with the uniform frequency solution $|\Omega|= \alpha v/\lambda$, the linear expression is valid up to $v \sim \alpha \tilde{A}Kt_p/2Ast_e$.